\documentclass[prb,nopacs,twocolumn,preprintnumbers,amsmath,amssymb]{revtex4}

\usepackage{epsfig}
\usepackage{graphicx}
\usepackage{dcolumn}
\usepackage{color}
\usepackage{bm}

\begin{document}

\title{Spin-Orbit mediated spin relaxation in monolayer MoS$_2$}

\author{H. Ochoa and R. Rold\'an}

{\affiliation{Instituto de Ciencia de Materiales de Madrid. CSIC. Sor Juana In\'es de la Cruz 3. 28049 Madrid. Spain.}

\begin{abstract}
We study the intra-valley spin-orbit mediated spin relaxation in monolayers of MoS$_2$ within a two bands effective Hamiltonian. The intrinsic spin splitting of the valence band as well as a Rashba-like coupling due to the breaking of the out-of-plane inversion symmetry are considered. We show that, in the hole doped regime, the out-of-plane spin relaxation is not very efficient since the spin splitting of the valence band tends to stabilize the spin polarization in this direction. We obtain spin lifetimes larger than nanoseconds, in agreement with recent valley polarization experiments.
\end{abstract}

\maketitle

\section{Introduction}

Among all the quasi two dimensional crystals that have become popular since the appearance of graphene,\citep{CastroNeto_etal} monolayer unit cells of molybdenum disulfide\citep{Mak_etal_2010,Splendiani_etal_2010} (MoS$_2$) and other dichalcogenides are particularly attractive because the existence of an electronic gap which makes those systems an excellent candidate for nanoelectronics devices.\citep{Radisavljevic_etal_2011,Zhang_etal,Wang_etal_2012,CR13} Whereas multilayer samples presents an indirect band gap, monolayer MoS$_2$ has a direct gap across the inequivalent K and K' points of the hexagonal Brilloin zone (BZ), which makes this material of particular interest for optoelectronic applications. Of special interest is the promising applications of MoS$_2$ for spintronics and valleytronics devices.\cite{Cao_etal_2012,Mak_etal_2012,Zeng_etal_2012}
In fact, the strong spin-orbit (SO) interaction, together with the absence of inversion symmetry in monolayer samples, splits the valence bands by $\sim 150$~meV in two spin flavors. This splitting is essential for spintronics and optoelectronics applications and, importantly, it has different signs at each valley due to time reversal symmetry. This fact  allows to control the valley population by optically exciting the monolayer samples with circularly polarized light, as it has been demonstrated experimentally,\citep{Mak_etal_2012,Zeng_etal_2012,Cao_etal_2012} what is generically called spin-valley coupling.\citep{Xiao_etal_2012}

A crucial role for the efficiency of the valley polarization is played by the spin lifetime $\tau_s$ of the system, which must be longer than $\sim 10$~ns for realistic applications. Interestingly, coherence times of this order have been experimentally measured for single layers of MoS$_2$.\cite{Mak_etal_2012} Therefore, understanding the spin relaxation mechanisms of this material is essential. Four mechanisms are usually discussed for spin relaxation in semiconductors:\citep{Zutic_etal} the Elliot-Yafet,\citep{Elliot,Yafet} D'yakonov-Perel',\citep{DyakonovPerel,Dyakonov} Bir-Aronov-Pikus,\citep{Bir_etal_1976} and hyperfine-interaction mechanisms.\citep{Dyakonov_Perel_1973} The latter, which accounts for the interaction between the magnetic moments of electrons and nuclei, is negligible in the diffusive regime due to the itinerant nature of the electrons. The hyperfine interaction with the nuclei spins is dynamically narrowed since the electrons move fast through  nuclei with random spins, averaging to zero their action. The Bir-Aronov-Pikus mechanism accounts for electron spin-flip processes mediated by the electron-hole exchange interaction, and it is typically relevant in heavily $p$-doped semiconductors.\citep{Bir_etal_1976,Song_Kim} The Elliot-Yafet and D'yakonov-Perel' mechanisms are mediated by the SO coupling. The former consists on the spin relaxation during a momentum scattering event by phonons or impurities, whereas the latter accounts for the spin precession in between scattering events induced by the SO coupling when inversion symmetry is broken.

Due to the strong SO coupling and the presence of disorder, both Elliot-Yafet and D'yakonov-Perel' mechanisms are expected to play a role, and they will be the main focus of the present work. Here we perform a systematic calculation of the intra-valley SO mediated spin relaxation rates by means of the Mori-Kawasaki formula,\citep{Mori_Kawasaki_1,Mori_Kawasaki_2} which is the appropriate framework to treat both mechanisms on the same footing.\citep{Boross_etal} Our results show that the intrinsic reflection symmetry of the system with respect to the out-of-plane direction in combination with the large spin splitting of the valence band allows spin lifetimes for the out-of-plane polarization larger than nanoseconds, in agreement with the experiments.\cite{Mak_etal_2012}

The manuscript is organized as follows. First, we present the two bands effective model that we employ to perform the calculation, including an exhaustive discussion on the microscopic origin of the intrinsic and Rashba-like SO couplings. In Sec.~\ref{sec:relaxation} we compute both the in-plane and out-of-plane spin relaxation rates associated to these couplings. Our results for SO mediated spin relaxation, as well as other alternative mechanism that may compete with them, are discussed in Sec. \ref{sec:discussion}. Finally, in Sec.~\ref{sec:conclusions} we summarize our main conclusions.

\section{The model}
\label{sec:model}
\subsection{Two bands effective model}

\begin{figure}
\begin{centering}
\includegraphics[width=0.49\columnwidth]{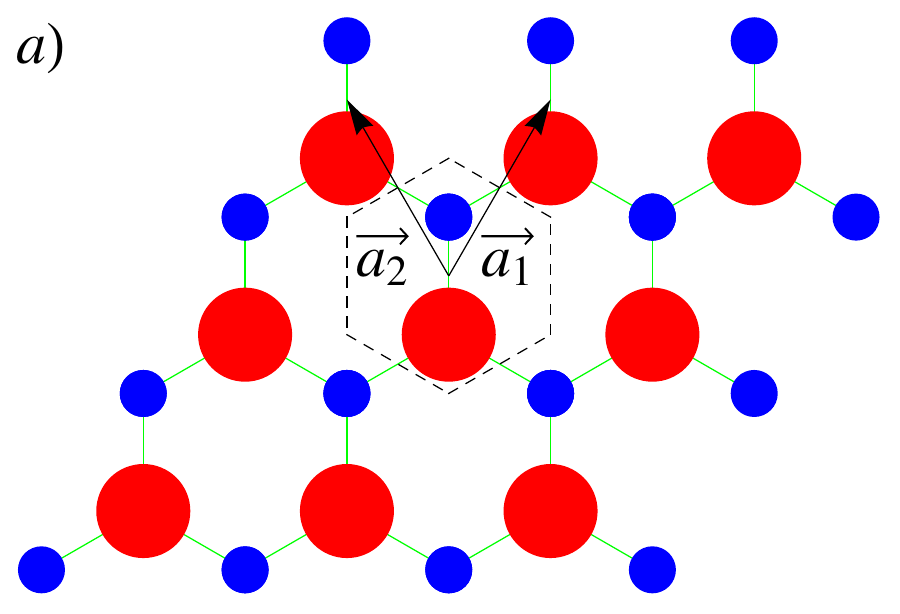}
\includegraphics[width=0.49\columnwidth]{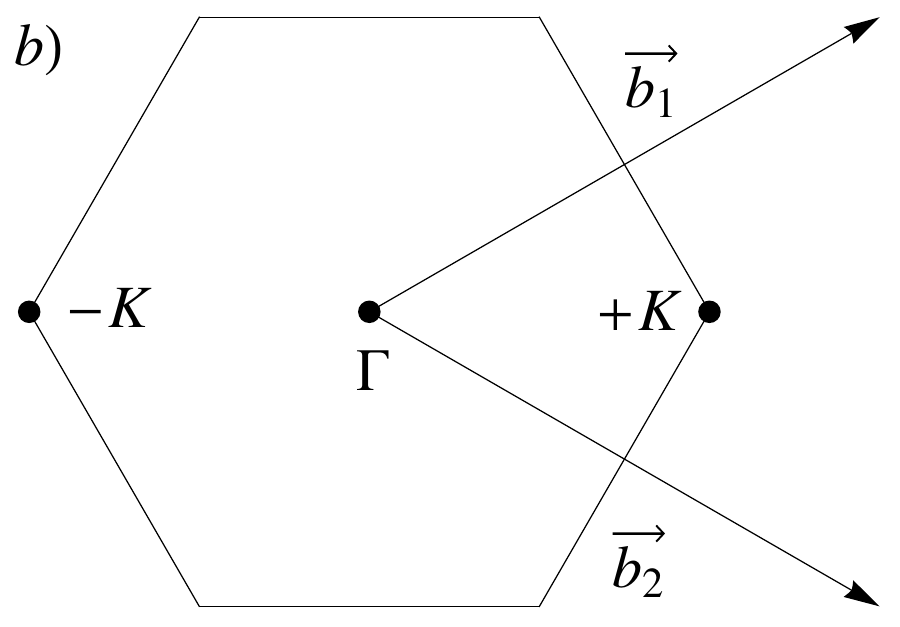}
\includegraphics[width=0.49\columnwidth]{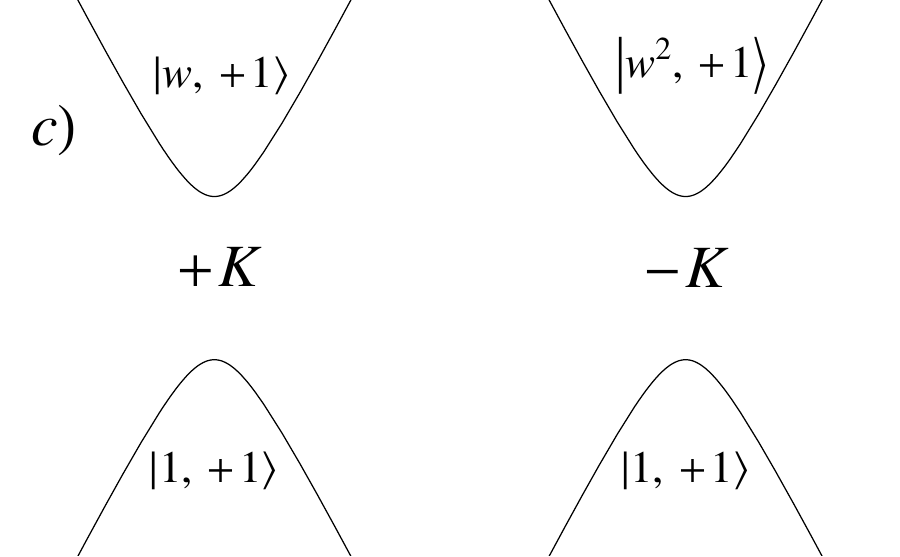}
\par\end{centering}
\caption{\label{Fig:lattice}a) Top view of the lattice in real space. Red and blue dots represent the Mo and S atoms, respectively. b) Brillouin zone, with the corresponding reciprocal lattice wave-vectors. The inequivalent $+\bf K$ and $-\bf K$  points are shown. c) Conduction and valence bands at $\pm\mathbf{K}$ points within the effective model discussed in the text. The SO splitting is not considered in this figure. The numbers inside the kets which label the bands express the phases picked up the Bloch wave function under symmetry operations of the crystal structure, see Tabs.~\ref{tab:character_tab} and \ref{tab:multiplets}.}
\end{figure}

The transition metal dichalcogenide MoS$_2$ is composed, in its bulk configuration, of two-dimensional S-Mo-S layers stacked on top of each other, coupled by weak van der Waals forces. The Mo atoms are ordered in a triangular lattice, each of them bonded to six S atoms located in the top and bottom layers, forming a sandwiched material. A top view of the lattice is shown in Fig. \ref{Fig:lattice} a). Like in graphene, the weak interlayer coupling makes possible to exfoliate this material down to a single-layer.\cite{NG05} The electronic band structure of MoS$_2$ changes from an indirect band gap for multilayer samples, to a direct gap semiconductor for single-layers, located at the two inequivalent $\pm {\bf K}$ points of the BZ [see Fig. \ref{Fig:lattice} b)].\cite{Splendiani_etal_2010} 

\begin{center}
\begin{table}
\begin{tabular}{|c|c|c|c|c|c|c|}
\hline
$C_{3h}=C_3\times\sigma_h$&$E$&$C_3$&$C_{3}^2$&$\sigma_h$&$S_3$&$\sigma_hC_3^2$\\
\hline
$A'$&1&1&1&1&1&1\\
\hline
$A''$&1&1&1&-1&-1&-1\\
\hline
$E'$&$\begin{array}{c}
1\\
1
\end{array}$&$\begin{array}{c}
w\\
w^2
\end{array}$&$\begin{array}{c}
w^2\\
w
\end{array}$&$\begin{array}{c}
1\\
1
\end{array}$&$\begin{array}{c}
w\\
w^2
\end{array}$&
$\begin{array}{c}
w^2\\
w
\end{array}$\\
\hline
$E''$&$\begin{array}{c}
1\\
1
\end{array}$&$\begin{array}{c}
w\\
w^2
\end{array}$&$\begin{array}{c}
w^2\\
w
\end{array}$&$\begin{array}{c}
-1\\
-1
\end{array}$&$\begin{array}{c}
-w\\
-w^2
\end{array}$&
$\begin{array}{c}
-w^2\\
-w
\end{array}$\\
\hline
\end{tabular}
\caption{Character table of $C_{3h}$. We denote $w=e^{\frac{i2\pi}{3}}$. Notation from Ref.~\onlinecite{Dresselhaus_book}.}
\label{tab:character_tab}
\end{table}
\end{center}

Being interested on single-layer samples, we compute here the spin relaxation rates within the two bands effective $\mathbf{k}\cdot\mathbf{p}$ Hamiltonian at the two inequivalent corners of the BZ. The hexagonal $D_{3h}$ symmetry of the monolayer crystal determines the explicit form of the Hamiltonian. Note that the $D_{3h}$ point group can be expressed as the direct product of $D_3$, which contains the identity, the two counterclockwise rotations by $2\pi/3$ and $4\pi/3$, and the reflections across the three in-plane axis which connect Mo and S atoms in a top view of the lattice [see Fig.~\ref{Fig:lattice} a)], and $\sigma_h$, which contains the identity and the inversion across the out-of-plane axis. The group of the wavevector at $\pm\mathbf{K}$ is $C_{3h}=C_3\times\sigma_h$ (the reflections across the in-plane axis swap the two inequivalent $\mathbf{K}$ points). The character table of this group can be found in Tab.~\ref{tab:character_tab}. Note that there is no two-dimensional irreducible representations of $C_{3h}$, so band touching at Dirac points are not protected by symmetry. Each band at $\pm\mathbf{K}$ can be labeled by two quantum numbers which express the phase picked up by the corresponding Bloch wave function at these points under a rotation by $2\pi/3$ and a reflection across the out-of-plane axis respectively, or equivalently, the irreducible representation of $C_{3h}$ associated to the Bloch wave function. These phases depend, of course, on the orbital character of these states. In Tab.~\ref{tab:multiplets} a classifiaction of the Bloch wave functions at the BZ corners for Mo orbitals ($s$, $p$, $d$) and S orbitals ($s$, $p$) is given.

Around $\pm\mathbf{K}$ points the conduction and valence bands are mostly made of $d$ orbitals coming from Mo. In particular, the orbital weight of the conduction band is essentially $d_{3z^2-r^2}$, which belongs to the $E'$ irreducible representation of $C_{3h}$, whereas in the case of the valence band is mostly the real combination of $d_{x^2-y^2}$ and $d_{xy}$ which belongs to $A'$. We restrict our analysis to a minimal low energy model that accounts for the above configuration. Up to first order in $\mathbf{k}$, the effective Hamiltonian reads:\citep{Xiao_etal_2012}
\begin{equation}
\mathcal{H}_0=at\left(\tau\sigma_xk_x+\sigma_yk_y\right)+\frac{\Delta}{2}\sigma_z
\label{eq:Hamiltonian}
\end{equation}where $t$ is the effective hopping amplitude, $a$ is the lattice constant, and the Pauli matrices $\sigma_i$ operate in a space of 2-component Bloch functions $\Psi_{\tau}=\left(\psi_{|w^{\tau},+1\rangle},\psi_{|1,+1\rangle}\right)^T$. Here $\left|w^{\tau}(1),+1(+1)\right\rangle$ labels the symmetry properties of the conduction (valence) band state, and $\tau=\pm 1$ corresponds to valley $\pm\mathbf{K}$.\footnote{A higher-order in $\bf k$ Hamiltonian, up to $k^2$, has been recently proposed by Rostani {\it et al.},\cite{RMA13} which includes different effective masses for the valence and conduction bands, as well as trigonal warping effects. However, for the low energy analysis of this work, Eq. (\ref{eq:Hamiltonian}) is a good approximation.}

\begin{center}
\begin{table}
\begin{tabular}{|c|c|c||c|c|}
\hline
Irreps&$C_3$&$\sigma_h$&Mo&S\\
\hline
\hline
$A'$&1&1&$\begin{array}{c}
\frac{1}{\sqrt{2}}\left(d_{x^2-y^2}\pm id_{xy}\right),\\
\frac{1}{\sqrt{2}}\left(p_x\mp ip_{y}\right)
\end{array}$
&$\frac{1}{\sqrt{2}}\left(p_x\pm ip_{y}\right)$ (s)\\
\hline
$A''$&1&-1&$\frac{1}{\sqrt{2}}\left(d_{xz}\mp id_{yz}\right)$&$\frac{1}{\sqrt{2}}\left(p_x\pm ip_{y}\right)$ (as)\\
\hline
$E'$&$w^{\pm1}$&1&$d_{3z^2-r^2}$, $s$&$\frac{1}{\sqrt{2}}\left(p_x\mp ip_{y}\right)$ (s)\\
\hline
$E'$&$w^{\mp1}$&1&
$\begin{array}{c}
\frac{1}{\sqrt{2}}\left(d_{x^2-y^2}\mp id_{xy}\right),\\
\frac{1}{\sqrt{2}}\left(p_x\pm ip_{y}\right)
\end{array}$
&$p_z$ (as), $s$ (as)\\
\hline
$E''$&$w^{\pm1}$&-1&$p_z$&$\frac{1}{\sqrt{2}}\left(p_x\mp ip_{y}\right)$ (as)\\
\hline
$E''$&$w^{\mp1}$&-1&$\frac{1}{\sqrt{2}}\left(d_{xz}\pm id_{yz}\right)$&$p_z$ (s), $s$ (s)\\
\hline
\end{tabular}
\caption{Classification of the Bloch wave functions at BZ corners according to the 6 irreducible representations of $C_{3h}$. The sign $\pm$ corresponds to $\pm\mathbf{K}$ points. In the case of S atoms, both symmetric (s) and anti-symmetric (as) combinations with respect to $\sigma_h$ of the orbitals of the top and the bottom atoms are considered. The second and third column contain the phases picked up by the wave function when a $2\pi/3$ rotation or a mirror reflection is performed.}
\label{tab:multiplets}
\end{table}
\end{center}

In order to discuss the effective SO coupling within this model we must introduce Pauli matrices associated to the spin degree of freedom. Importantly, we are now introducing a pseudovector in the 3-dimensional space, meaning that the operators which contain $s_z$ are even under $\sigma_h$ whereas the operators which contain the in-plane component of the spin are odd. Unless the $\sigma_h$ symmetry is expressly broken, our effective model may only contain terms with $s_z$. Therefore, the $\sigma_h$ symmetry of the crystal structure protects the out-of-plane spin component. Then, the intrinsic SO coupling terms read in general:
\begin{equation}
\mathcal{H}_{int}^{SO}=\lambda_{c}\tau\frac{\mathcal{I}+\sigma_z}{2}s_z+\lambda_v\tau\frac{\mathcal{I}-\sigma_z}{2}s_z
\end{equation}
where $\cal I$ is the identity matrix acting in the space of 2-component Bloch functions. The absence of a center of inversion in the crystal implies the spin splitting of the bands. Here $\lambda_{c}$ and $\lambda_v$ are the splittings of the conduction and valence bands respectively. Although both splittings are allowed by symmetry, it is important to notice their different microscopic origins due to the different orbital character of the bands. We consider an intra-atomic SO Hamiltonian of the form $\Delta_{SO}\mathbf{L}\cdot\mathbf{s}$ for the $d$ orbitals of Mo, where $\mathbf{L}$ is the angular momentum operator. As we show schematically in Fig.~\ref{fig:SOC}, the spin splitting of the valence band is the result of a first order process. However, the splitting of the conduction band is associated to second order processes which involve virtual transitions into states wich belong to the $A''$ and $E''$ irreducible representations. Therefore, 
\begin{equation}
\lambda_{int}\equiv\lambda_v\gg\lambda_c,
\end{equation}
and we neglect from here on the SO splitting of the conduction band.

The values of the model parameters can be extracted from experiments as well as from first principle calculations.\citep{M73,Coehoorn_etal,Kuc_etal_2011,Zhu_etal_2011,Feng_etal_2012}
Here we take $at=3.51$~eV\AA, $\Delta=1.66$ eV and $2\lambda_{int}=0.15$ eV, with the lattice constant $a=3.193$ \AA.

\subsection{Rashba effect}

\begin{figure}
\begin{centering}
\includegraphics[width=1\columnwidth]{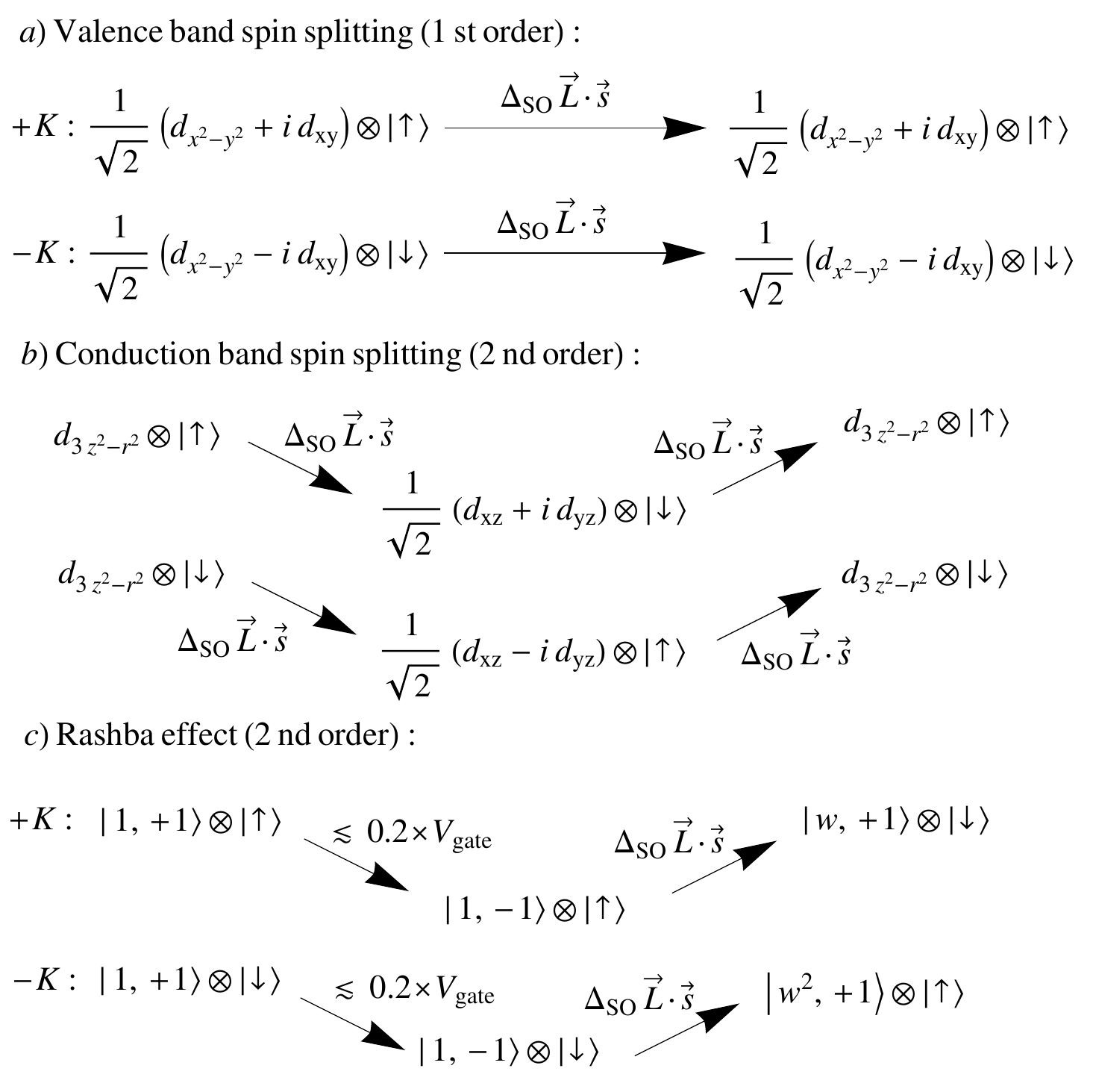}
\par\end{centering}
\caption{\label{fig:SOC}Sketch of the microscopic processes which lead to the effective SO coupling terms discussed in the text. a) First order processes which lead to the splitting of the valence band. b) Second order processes associated to the splitting of the conduction band. c) Second order processes which lead to a Bychkov-Rashba coupling when $\sigma_h$ symmetry is broken.}
\end{figure}

If the $\sigma_h$ symmetry is broken then $s_z$ is not longer a good quantum number. In that situation a coupling with the in-plane components is possible. A Bychkov-Rashba coupling\citep{Bychkov_Rashba} appears, which in the context of the two bands effective model reads:\begin{equation}
\mathcal{H}_{ext}^{SO}=\lambda_{ext}\left(\tau\sigma_xs_y-\sigma_ys_x\right).
\end{equation}
This coupling is the result of second order processes as in the case of the spin splitting of the conduction band, although first order in the SO interaction as it is shown in Fig.~\ref{fig:SOC} c).

The Rashba effect in 2D systems is usually attributed to dipolar transitions induced by the aplication of an electric field in the out-of-plane direction. In our case, such field would induce dipolar transitions between the Mo $d_{3z^2-r^2}$ orbitals of the conduction band and Mo $p_z$ orbitals of bands at much higher energies. Then, the SO interaction would induce transitions between these states and the valence band flipping the spin. However, the orbital weight of the valence band in Mo $p$ orbitals is very small, so this coupling is expected to be very weak.
Nevertheless, this picture changes if orbitals from S atoms are also taken into account. For instance, if we consider the application of a gate voltage $V_{gate}$, which is necessary in order to induce charge carriers in this system, then we would have different on-site energies for the $p$ orbitals of the top and bottom S atoms. This turns into a non zero hybridization
between valence band states $\left|1, +1\right\rangle$ and states $\left|1, -1\right\rangle$ of higher bands proportional to $V_{gate}$. Then, the SO interaction induce transitions between $d_{xz}$, $d_{yz}$ orbitals of these bands and $d_{3z^2-r^2}$ of the conduction band, flipping the spin. This kind of processes is the one depicted in Fig.~\ref{fig:SOC} c). Since the orbital weigth of S $p$ orbitals in these bands is less than the 20 $\%$, we can estimate an upper limit for this coupling of the form:\begin{align}
\lambda_{ext}\leq\frac{0.2V_{gate}\Delta_{SO}}{\epsilon_{\left|1, -1\right\rangle}}
\end{align}
where $\epsilon_{\left|1, -1\right\rangle}$ represents the energy of the $\left|1, -1\right\rangle$ ($A''$) band involved in the calculation.

\begin{figure}
\begin{centering}
\includegraphics[width=1\columnwidth]{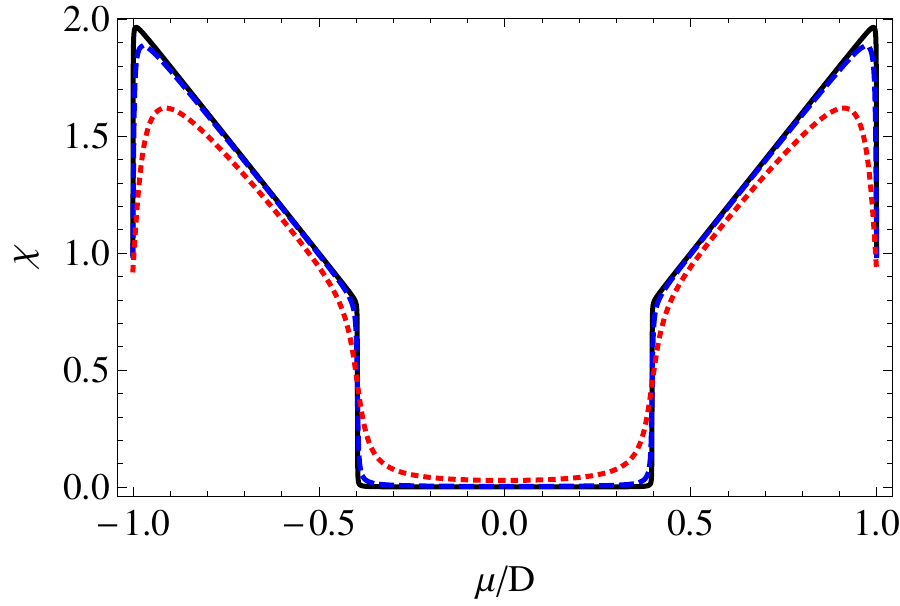}
\par\end{centering}
\caption{\label{fig:chi}Spin susceptibility $\chi$ as a function of the chemical potential $\mu$, for different values of $\Gamma$, calculated from Eq. (\ref{eq:chi}). Solid black line corresponds to $\Gamma=0.001$ eV, blue dashed line to $\Gamma=0.01$ eV, and red dotted line to $\Gamma=0.1$ eV.}
\end{figure}

\subsection{Disorder and Mori-Kawasaki formula}

We compute the spin relaxation rates by using the Mori-Kawasaki formula.\citep{Mori_Kawasaki_1,Mori_Kawasaki_2} Originally, the Mori-Kawasaki formula was deduced in order to compute the broadening of the signal peak in an electron spin resonance experiment due to the breaking of the $SU(2)$ spin symmetry of the system. As it has been shown recently,\citep{Boross_etal} it can be related with the inverse of the spin lifetime $\Gamma_s=\hbar/\tau_s$. The Mori-Kawasaki formula treats the SO coupling terms as perturbations to the electronic Hamiltonian, something that in principle is valid since in our model $\lambda_{int}$, $\lambda_{ext}$ are at least one order of magnitude smaller than the other energy parameters of the model, namely the gap $\Delta$ and the bandwidth $2t$. The spin lifetime is computed as $\Gamma_s$:
\begin{equation}
\Gamma_s=-\frac{1}{\chi} \lim_{\omega\rightarrow 0}{\rm Im}\frac{\chi_{\mathcal{A}\mathcal{A}^{\dagger}}\left(\omega\right)}{\omega}
\label{eq:Mori-Kawasaki}
\end{equation}
where $\chi$ is the spin susceptibility and
$\chi_{\mathcal{A}\mathcal{A}^{\dagger}}\left(\omega\right)$ is the Fourier transform of the response function:\begin{equation}
\chi_{\mathcal{A}\mathcal{A}^{\dagger}}(t)=-i\theta(t)\left\langle\left[\mathcal{A}(t),\mathcal{A}^{\dagger}(0)\right]\right\rangle
\label{eq:definitionX}
\end{equation}
with $\mathcal{A}=\left[\mathcal{H}^{SO},s_+\right]$ and $s_{\pm}=(s_x\pm is_y)/2$. The spin susceptibility is defined as:
\begin{equation}\label{eq:chi_def}
\chi=\frac{1}{g\mu_B}\left(\frac{\partial \langle s_z\rangle}{\partial H}\right)_{H=0}
\end{equation}
where $H$ is the field strength associated to a Zeeman term in the Hamiltonian ${\cal H}_Z=-g\mu_BHs_z$.\cite{OA02}
The expectation values in Eqs.~\eqref{eq:definitionX} and \eqref{eq:chi_def} are referred to the Hamiltonian without the SO coupling terms.

For SO mediated spin relaxation mechanisms, particularly the Elliot-Yafet\citep{Elliot,Yafet} and the D'yakonov-Perel' mechanisms,\citep{DyakonovPerel,Dyakonov} a relevant parameter of the theory is the amount of disorder $\Gamma=\hbar/\tau_{\mathbf{p}}$, where $\tau_{\mathbf{p}}$ is the lifetime of the quasiparticles with momentum $\mathbf{p}$. Disorder is introduced in our model in a phenomenological manner, by adding the imaginary self-energy $i\Gamma/2$ to the Matsubara Green's operator associated to the free Hamiltonian~\eqref{eq:Hamiltonian}:
\begin{equation}
\hat{ G}(\mathbf{k},i\omega)=\frac{1}{2}\sum_{\alpha=\pm1}G_{\alpha}(\mathbf{k},i\omega)\left[\mathcal{I}+\alpha\left(at\mathbf{k}\cdot\vec{\sigma}+\frac{\Delta}{2}\sigma_z\right)/\epsilon_{\mathbf{k}}\right]
\end{equation}
where we define:
\begin{equation}
\label{eq:green}
G_{\alpha}(\mathbf{k},i\omega)=\frac{1}{i\omega-\alpha\epsilon_{\mathbf{k}}+\mu+i\Gamma/2}.
\end{equation}
Here $\epsilon_{\mathbf{k}}=\sqrt{a^2t^2|\mathbf{k}|^2+\Delta^2/4}$ is the dispersion relation of conduction ($\alpha=+1$) and valence ($\alpha=-1$) bands of our effective model and $\mu$ is the chemical potential. Note that the valley index $\tau$ is omitted since we do not consider short-range scatterers which could connect both valleys. Therefore, inter-valley relaxation processes are beyond the scope of this work. We note that the inclusion of inter-valley disorder in combination with the SO interaction and a possible $\sigma_h$ symmetry breaking can lead to interesting localization phenomena.\citep{Falko,Lu_etal_2013}}
We also neglect the momentum dependence of $\Gamma$, so it enters just as a parameter which in principle can be determined from transport experiments.\citep{Zhang_etal} 

The expectation value of the $z$-component of spin is:
\begin{equation}
\langle s_z \rangle = \frac{1}{N}\sum_{{\bf k}}(n_{{\bf k}\uparrow}-n_{{\bf k}\downarrow})
\end{equation}
where $N$ is the number of unit cells and $n_{{\bf k}s}$ is the occupation number of quasiparticles with momentum $\mathbf{k}$ and spin $s$ in the presence of the Zeeman term ${\cal H}_Z$. This can be calculated in terms of the spectral functions $A_{\alpha}({\bf k},\omega)$, defined from the retarded version of the Green's functions of Eq.~\eqref{eq:green} as:
\begin{equation}
A_{\alpha}\left(\mathbf{k},\omega\right)\equiv-2~{\rm Im}G_{\alpha}^R(\omega,\mathbf{k})=\frac{\Gamma}{\left(\omega-\alpha\epsilon_{\mathbf{k}}+\mu\right)^2+\frac{\Gamma^2}{4}}.
\end{equation}
Then, we can write:
\begin{equation}
n_{{\bf k}s}=2\sum_{\alpha=\pm1}\int_{-\infty}^{\infty}\frac{d\omega}{2\pi}n_F(\omega)A_{\alpha}({\bf k},\omega+sg\mu_BH).
\end{equation}
where $n_F(\omega)$ is the Fermi-Dirac distribution function and the factor 2 accounts for the valley degeneracy. From the definition of Eq.~\eqref{eq:chi_def} we get:\begin{equation}
\chi=4\sum_{\alpha=\pm 1}\frac{1}{N}\sum_{\mathbf{k}}\int_{-\infty}^{\infty}\frac{d\omega}{2\pi}A_{\alpha}({\bf k},\omega)\left(-\frac{\partial n_F(\omega)}{\partial \omega}\right)
\end{equation}
In the zero temperature limit ($T\ll T_F$, where $T_F$ is the Fermi temperature) we can approximate $-\frac{\partial n_F(\omega)}{\partial \omega}\approx\delta(\omega)$. The sum in $\mathbf{k}$ can be written as an integral through the standard procedure $\frac{1}{N}\sum_{\mathbf{k}}\rightarrow\frac{A_c}{(2\pi)^2}\int d^2\mathbf{k}$, where $A_c$ is the area of the unit cell. The isotropy of the dispersion relation allows to integrate in angles straightforwardly and to write down the remaining integral in $|\mathbf{k}|$ as an integral in energies. At this point, it is necessary to introduce an energy cutoff $D$ for the effective model, which can be related with the area of the unit cell as $D=at\sqrt{\pi/A_c}$. After some algebra the spin susceptibility can be written as:
\begin{eqnarray}
\chi&=&\frac{\Gamma A_c}{\pi^2 a^2t^2}\sum_{\alpha=\pm 1}\int_{\frac{\Delta}{2D}}^1dx\frac{x}{\left(x-\alpha\frac{\mu}{D}\right)^2+\frac{\Gamma^2}{4D^2}}
\label{eq:chi}
\end{eqnarray}
The spin susceptibilty as a function of the chemical potential is shown, for different values of $\Gamma$, in Fig.~\ref{fig:chi}. If we drop logarithmically small terms that appear  in Eq.~\eqref{eq:chi} after integration in $x$, we obtain a simple analytical formula for $\chi$ which is valid for $\mu\ll D$:
\begin{align}
\label{eq:chi_anal}
\chi\approx\frac{2\mu A_c}{\pi^2a^2t^2}\sum_{\alpha=\pm1}\arctan\left(\frac{2\mu-\alpha\Delta}{\Gamma}\right).
\end{align}

We compute now the numerator of Eq.~\eqref{eq:Mori-Kawasaki}. The calculation is easily performed in the Matsubara frequency domain. We can write:
\begin{equation}
\chi_{\mathcal{A}\mathcal{A}^{\dagger}}(i\omega)=\frac{1}{\beta N}\sum_{\mathbf{k}}\sum_{i\nu}\sum_{\alpha,\alpha'}f_{\alpha\alpha'}\left(\mathbf{k}\right)G_{\alpha}\left(\mathbf{k},i\omega+i\nu\right)G_{\alpha'}\left(\mathbf{k},i\nu\right)
\end{equation}
where $\beta$ is the usual thermal factor and $f_{\alpha\alpha'}\left(\mathbf{k}\right)$ is defined as:
\begin{widetext}\begin{equation}
f_{\alpha\alpha'}\left(\mathbf{k}\right)=\frac{1}{2}{\rm Tr}\left[\mathcal{A}\cdot\left(\mathcal{I}+\alpha\frac{at\mathbf{k}\cdot\vec{\sigma}+\frac{\Delta}{2}\sigma_z}{\epsilon_{\mathbf{k}}}\right)\cdot\mathcal{A}^{\dagger}\cdot\left(\mathcal{I}+\alpha'\frac{at\mathbf{k}\cdot\vec{\sigma}+\frac{\Delta}{2}\sigma_z}{\epsilon_{\mathbf{k}}}\right)\right]
\end{equation}\end{widetext}
The trace is performed in the space of 2-components Bloch functions, and the valley degeneracy has already been taken into account in this definition. The sum in frequencies can be performed easly by using the Lehmann representation in terms of the spectral functions introduced before. After the sumation and the analytical continuation we have for the imaginary part of $\chi_{\mathcal{A}\mathcal{A}^{\dagger}}\left(\omega\right)$:

\begin{align}
-{\rm Im}\chi_{\mathcal{A}\mathcal{A}^{\dagger}}\left(\omega\right)=\frac{1}{N}\sum_{\mathbf{k}}\sum_{\alpha,\alpha'}f_{\alpha\alpha'}\left(\mathbf{k}\right)\times
\nonumber\\
\times\int_{-\infty}^{\infty}\frac{d\epsilon}{4\pi}A_{\alpha}\left(\mathbf{k},\epsilon+\omega\right)A_{\alpha'}\left(\mathbf{k},\epsilon\right)\left[n_F\left(\epsilon\right)-n_F\left(\epsilon+\omega\right)\right]
\end{align}
Hence, in the $\omega\rightarrow0$ limit we obtain:
\begin{align}
\lim_{\omega\rightarrow 0}-{\rm Im}\frac{\chi_{\mathcal{A}\mathcal{A}^{\dagger}}\left(\omega\right)}{\omega}=\frac{1}{N}\sum_{\mathbf{k}}\sum_{\alpha,\alpha'}f_{\alpha\alpha'}\left(\mathbf{k}\right)\times
\nonumber\\
\times\int_{-\infty}^{\infty}\frac{d\epsilon}{4\pi}A_{\alpha}\left(\mathbf{k},\epsilon\right)A_{\alpha'}\left(\mathbf{k},\epsilon\right)\left(-\frac{\partial n_F\left(\epsilon\right)}{\partial \epsilon}\right)
\end{align}
After the same approximations as before we can write, in the zero temperature limit:
\begin{equation}
\Gamma_s=\frac{1}{4\pi\chi}\frac{A_c}{(2\pi)^2}\sum_{\alpha,\alpha'}\int d^2\mathbf{k}f_{\alpha\alpha'}\left(\mathbf{k}\right)A_{\alpha}\left(\mathbf{k},0\right)A_{\alpha'}\left(\mathbf{k},0\right)
\label{eq:gamma_s}
\end{equation}
The remaining part of the paper will be devoted to the estimation, using Eq. (\ref{eq:gamma_s}), of the spin relaxation in the different scenarios which are relevant for MoS$_2$.

\section{Spin relaxation}
\label{sec:relaxation}

\subsection{In-plane spin relaxation}

We start by computing the in-plane spin relaxation rate due to the intrinsic SO coupling. For this, we use Eq. \eqref{eq:gamma_s} with $\mathcal{A}=\lambda_{int}\left(\mathcal{I}-\sigma_z\right)s_+$, which leads to:
\begin{equation}
f_{\alpha\alpha'}\left(\mathbf{k}\right)=2\lambda_{int}^2\left(1-(\alpha+\alpha')\frac{\Delta}{2\epsilon_{\mathbf{k}}}+\alpha\alpha'\frac{\Delta^2}{4\epsilon_{\mathbf{k}}^2}\right)
\end{equation}
Then, one sees that the in-plane relaxation rate can be written as the sum of two contributions, one coming from intra-band transitions and the other from inter-band transitions:
\begin{equation}
\Gamma_{in}=\frac{\lambda_{int}^2\Gamma^2A_c}{2\pi^2\chi a^2t^2 D^2}\left[I_{intra}+I_{inter}\right]
\end{equation}
where:
\begin{eqnarray}
I_{intra}&=&\frac{1}{2}\sum_{\alpha=\pm 1}\int_{\frac{\Delta}{2D}}^1dx\frac{x-\alpha\frac{\Delta}{D}+\frac{\Delta^2}{4xD^2}}{\left[\left(x-\alpha\mu\right)^2+\frac{\Gamma^2}{4}\right]\left[\left(x-\alpha\mu\right)^2+\frac{\Gamma^2}{4}\right]}\nonumber\\
I_{inter}&=&\int_{\frac{\Delta}{2D}}^1dx\frac{x-\frac{\Delta^2}{4xD^2}}{\left[\left(x-\mu\right)^2+\frac{\Gamma^2}{4}\right]\left[\left(x+\mu\right)^2+\frac{\Gamma^2}{4}\right]}
\end{eqnarray}

\begin{figure}
\begin{centering}
\includegraphics[width=1\columnwidth]{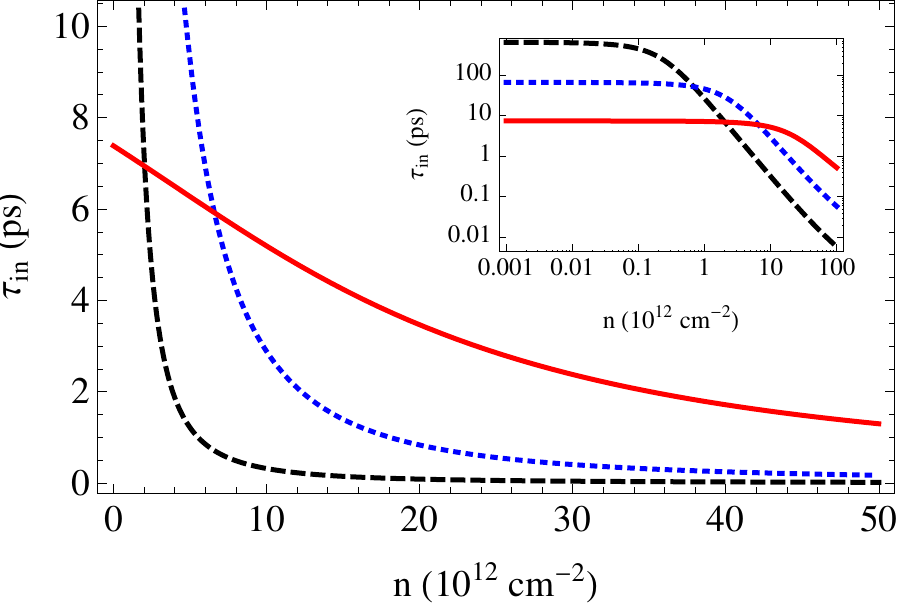}
\includegraphics[width=1\columnwidth]{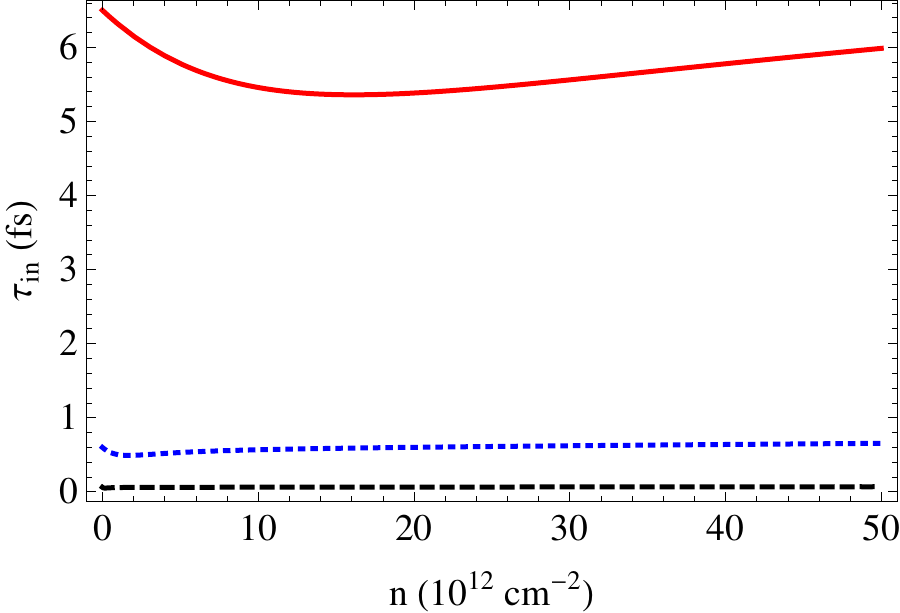}
\par\end{centering}
\caption{\label{fig:in-plane}In-plane spin lifetimes as a function of the carrier concentration. Top: Electron doping. Bottom: Hole doping. Dashed black line corresponds to $\Gamma=0.001$ eV, dotted blue to $\Gamma=0.01$ eV, and solid red $\Gamma=0.1$ eV. Inset: In-plane spin lifetimes for electron concentrations in double logarithmic scale. Notice the different time scale in the top and bottom panels.}
\end{figure}

The intra-band transitions account the D'yakonov-Perel' processes, whereas the inter-band term leads to the Elliot-Yafet contribution. This is more clear in the doped regime $|\mu|>\Delta/2$. If we drop logarithmic corrections in the above integrals, as we did in order to get Eq.~\eqref{eq:chi_anal}, we arrive at:\begin{align}
\Gamma_{in}^{intra}\approx\frac{\lambda_{int}^2}{2\Gamma}\left[1-\frac{\Delta}{\mu}+\frac{\Delta^2\left(\mu^2+\frac{3\Gamma^2}{4}\right)}{4\left(\mu^2+\frac{\Gamma^2}{4}\right)^2}\right]\nonumber\\
\Gamma_{in}^{inter}\approx\frac{\lambda_{int}^2\Gamma}{8\mu^2}\left[1-\frac{\Delta^2\left(\mu^2-\frac{\Gamma^2}{4}\right)}{4\left(\mu^2+\frac{\Gamma^2}{4}\right)^2}\right]
\end{align}
The inter-band transitions lead to an Elliot-Yafet contribution characterized by the linear scaling between the spin lifetime and momentum scattering time $\Gamma_{in}^{inter}\propto\Gamma$. The intra-band transitions, however, leads to the D'yakonov-Perel' mechanism, characterized by $\Gamma_{in}^{intra}\propto\Gamma^{-1}$. This mechanism is clearly the dominant one, as expected from symmetry considerations, due to the absence of a center of inversion in the crystal structure. Assuming that $\Gamma\ll\mu,\Delta$ we have:\begin{equation}
\frac{\Gamma_{in}^{intra}}{\Gamma_{in}^{inter}}\approx\frac{1-\frac{\Delta}{2\mu}}{1+\frac{\Delta}{2\mu}}\cdot\left(\frac{\mu}{\Gamma/2}\right)^2
\end{equation}and therefore $\Gamma_{in}^{intra}/\Gamma_{in}^{inter}\gg1$ unless the chemical potential lies at the bottom of the conduction band.
It is important to note that the D'yakonov-Perel' mechanism is clearly electron-hole asymmetric due to the different spin splittings of the conduction and valence bands.

These features are clearly shown in Fig.~\ref{fig:in-plane}, where the in-plane spin lifetime is computed numerically. We see that the D'yakonov-Perel' mechanism is clearly dominant for hole dopings. From mobilities reported in transport experiments\citep{Zhang_etal} we deduce $\Gamma\approx 0.02$ eV, and therefore $\tau_{in}\sim2\hbar\Gamma/\lambda_{int}^2\approx 5$ fs. For electron concentrations, it is interesting to note the crossover from D'yakonov-Perel' to Elliot-Yafet dominated regimes when the concentration is decreased, as it can be seen in the inset of the top panel of Fig.~\ref{fig:in-plane}. Such crossover is possible when the strength of disorder is comparable with the chemical potential measured with respect to the bottom of the band. In this case the spin lifetimes are 3 orders of magnitude larger than in the case of hole doping. Note that in the electron doped regime a more realistic calculation should take into account also the spin splitting of the conduction band.

\subsection{Out-of-plane spin relaxation}

We compute now the out-of-plane spin relaxation rate due to an extrinsic or Rashba-like coupling. In this case we have $\mathcal{A}=-2i\lambda_{ext}\sigma_+s_z$, which leads to:
\begin{equation}
f_{\alpha\alpha'}\left(\mathbf{k}\right)=4\lambda_{ext}^2\left(1-(\alpha-\alpha')\frac{\Delta}{2\epsilon_{\mathbf{k}}}-\alpha\alpha'\frac{\Delta^2}{4\epsilon_{\mathbf{k}}^2}\right)
\end{equation}
The calculation is formally identical to the previous one. In the doped regime we have the approximate results:\begin{align}
\Gamma_{out}^{intra}\approx\frac{\lambda_{ext}^2}{\Gamma}\left[1-\frac{\Delta^2\left(\mu^2+\frac{3\Gamma^2}{4}\right)}{4\left(\mu^2+\frac{\Gamma^2}{4}\right)^2}\right]\nonumber\\
\Gamma_{out}^{inter}\approx\frac{\lambda_{ext}^2\Gamma}{4\mu^2}\left[1+\frac{\Delta^2\left(\mu^2-\frac{\Gamma^2}{4}\right)}{4\left(\mu^2+\frac{\Gamma^2}{4}\right)^2}\right]
\label{eq:gamma_out}
\end{align}

\begin{figure}
\begin{centering}
\includegraphics[width=1\columnwidth]{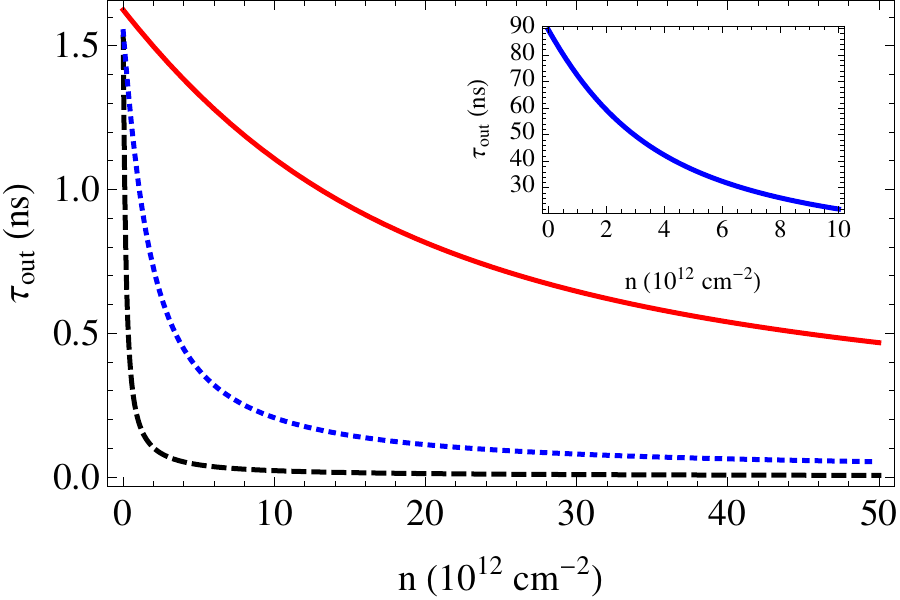}
\par\end{centering}
\caption{\label{fig:out-of-plane}Out-of-plane spin lifetimes as a function of the carrier concentration. In black (dashed) $\Gamma=0.001$ eV, in blue (dotted) $\Gamma=0.01$ eV, in red $\Gamma=0.1$ eV. In all the cases $\lambda_{ext}=10^{-2}\lambda_{int}$. Inset: Spin lifetime for hole concentrations where the correction given by Eq.~\eqref{eq:correction}.}
\end{figure}

In Fig.~\ref{fig:out-of-plane} the numerical computation of the spin lifetime as a function  of the carrier concentration is shown. We take $\lambda_{ext}=10^{-2}\lambda_{int}$, which is the correct order of magnitude given that this coupling is the result of second order processes as we explained in the previous section. The spin lifetimes are in this case electron-hole symmetric and clearly dominated by the D'yakonov-Perel' mechanism. The $1/n$ behavior is expected from the first expression in Eq.~\eqref{eq:gamma_out}. For $\mu\geq\Delta/2$ we have $\tau_{out}\sim\hbar\Gamma\Delta^2/(4\pi \lambda_{ext}^2a^2t^2n)$, so for $\Gamma=0.02$ eV and $n=10^{12}$ cm$^{-2}$ we obtain $\tau_{out}\approx 1-2$ ns.

Note that the spin splitting of the valence band is not taken into account in this calculation, but its effect is relevant since it tends to stabilize the out-of-plane spin polarization, in a similar way as an applied magnetic field in the out-of-plane direction does. We can take into account this effect by correcting the spin relaxation rate for hole concentrations as:\citep{Dyakonov,Boross_etal}\begin{equation}
\Gamma_{out}^{(holes)}\approx\Gamma_{out}\times\frac{1}{1+\left(\frac{2\lambda_{int}}{\Gamma}\right)^2}
\label{eq:correction}
\end{equation}where $2\lambda_{int}$ can be interpreted as the Zeeman splitting created by an effective magnetic field whose origin is the intrinsic SO coupling. Therefore, for $\Gamma=0.02$ eV and $n=10^{12}$ cm$^{-2}$ we expect:\begin{equation}
\tau_{out}^{(holes)}\approx60\times\tau_{out}\approx50-100\mbox{ ns}
\end{equation}
This correction is taken into account in the calculation shown in the inset of Fig.~\ref{fig:out-of-plane}. It is interesting to notice that our results quantitatively agree with the experimental measurements of Mak {\it et al.}, who have reported spin lifetimes exceeding 1ns in single layers of MoS$_2$.\cite{Mak_etal_2012}

\section{Discussion}
\label{sec:discussion}

The $\sigma_h$ symmetry preserves the out-of-plane spin polarization. However, in real systems this symmetry is broken by the presence of a substrate, electric fields, ripples, etc. Our calculation shows that, for a realistic value of the extrinsic or Rashba-like coupling generated by this symmetry breaking, the spin lifetimes are of the order of nanoseconds. Moreover, the large splitting of the valence band due to the spin-obit coupling contributes to stabilize the out-of-plane spin polarization, as a magnetic field in that direction does, and therefore the spin lifetimes in the hole doped regime are expected to be at least one order of magnitude larger, in agreement with recent experiments.\citep{Mak_etal_2012,Zeng_etal_2012,Cao_etal_2012}

Although our previous results suggest that the SO mediated D'yakonov-Perel' mechanism can account for the spin relaxation in single layers of MoS$_2$, another (non SO mediated) mechanism could be operative in this material. As proposed in Ref. \onlinecite{Mak_etal_2012}, the Bir-Aronov-Pikus mechanism\cite{Bir_etal_1976} can be efficient in $n$-doped samples of MoS$_2$, where the spin of a conduction electron may be flipped by the exchange interaction with a hole. The hole intervening in such a process was identified in Ref. \onlinecite{Mak_etal_2012} with a negative trion. In fact, the two-dimensionality of this system leads to an enhancement of the Coulomb interaction, due to the reduced dielectric screening, which can favor the formation of tightly bounded trions. Such quasiparticles, which have been recently measured experimentally,\citep{Mak_etal_2013} are formed by an exciton (electron-hole pair) bounded with an extra electron, therefore carrying a negative charge. In this situation, a  mechanism for spin relaxation analogue to the conventional Bir-Aronov-Pikus for heavily $p$-doped semiconductors, would be possible but with a trion playing here the role of the exchanged hole.\cite{Mak_etal_2012} 

An estimation of the efficiency of such a relaxation process  would require the knowledge of some parameters, as the trion effective mass, the exchange splitting of the excitonic ground state, the electron density induced by unintentional $n$-doping, or the band velocities and the Sommerfeld's factor, which determines the electron-hole overlap amplitude and which depends on the strength of the Coulomb interaction.\cite{Bir_etal_1976} In the absence of an accurate determination of the above quantities, the approximations used in Ref. \onlinecite{Mak_etal_2012} suggest that this mechanism might be relevant, leading to relaxation times of the order of nanoseconds.

Furthermore, it is worth to mention that the $\sigma_h$ symmetry is intrinsically broken by the out-of-plane (flexural) phonon modes of the MoS$_2$ monolayer, which constitutes an additional source of out-of-plane spin relaxation,\citep{Song_Dery} as it happens in graphene.\citep{Ochoa_etal_2012,Fratini_etal} In the free-standing system the dispersion of the flexural acoustic branch is quadratic,\citep{Molina-Sanchez_Wirtz_2011} as expected from symmetry considerations.
The bending rigidity of the system, which determines the energy of these modes, can be estimated assuming a simplified model where the MoS$_2$ sheet is described as a plate of certain thickness $\delta$. The bending rigidity reads:\citep{Landau_elasticity}
\begin{equation}
\kappa=\frac{Y\delta^3}{24\left(1-\sigma^2\right)}
\end{equation}where $Y=0.33$ TPa  is the Young modulus\citep{Castellanos_etal_2012} and $\sigma=0.125$ is the Poisson ratio.\citep{Lovell_etal} If we take as $\delta$ the inter-layer distance, $\delta\approx6.75$ \AA,\citep{Benamen_etal_2011} then we obtain $\kappa\approx27$ eV. Although it is difficult to judge the accuracy of this estimation due to the $\delta^3$ dependence, it is reasonable to take a bending rigidity bigger than in graphene, even an order of magnitude.\citep{Schwarz_etal} Therefore, one expects that the higher stiffness of MoS$_2$ and the spin splitting of the bands will protect the out-of-plane spin polarization.

Regarding the in-plane spin relaxation, it is clear that our results for electron doping are limited by the fact that the splitting of the conduction band is not included in our calculations ($\lambda_c\sim$ 3 meV according to recent estimations\citep{Kormanyos_etal_2013}). Nevertheless, the splitting of the valence band provides a remarkable source of relaxation. This is so because the conduction and valence band states are strongly hybridized away from $\pm\mathbf{K}$ points (note that $t\sim$ 1 eV, similar to the gap $\Delta$), which actually justifies the use of the two bands model. As a consequence, it turns out that the use of the two bands model is essential in order to take into account both the Elliot-Yafet and the D'yakonov-Perel' mechanisms in the conduction band. Very recently, a single band model has been proposed in order to explain in-plane spin relaxation in MoS$_2$ monolayer.\citep{Wang_Wu_2013} In that case, the splitting of the band together with the intervalley electron-phonon scattering opens an intervalley spin relaxation channel which may compete with the intra-valley one discussed here.

\section{Conclusions}
\label{sec:conclusions}

We have computed the spin lifetimes of monolayer MoS$_2$ within a two bands effective model. Assuming an extrinsic Rashba-like coupling generated by a $\sigma_h$ symmetry breaking of the order of $\lambda_{ext}\sim 10^{-2}\lambda_{int}$, we have  obtained spin lifetimes of the order of $\tau_{out}\sim100$ ns, estimation which is in agreement with recent valley population experiments.\cite{Mak_etal_2012}

Our calculations show that the D'yakonov-Perel' mechanism dominates the SO mediated spin relaxation in monolayers of MoS$_2$. The method used here is completely general, and the results can be extrapolated to other dichalcogenides by simply replacing the values of the model parameters. Of special interest is the case of WS$_2$ monolayers, for which even longer relaxation times are expected since it presents an even larger SO coupling, of the order of $\sim 400$ meV.\cite{ZC12} In general, the in-plane spin relaxation is very efficient due to the strong SO interaction and the lack of inversion symmetry of the system. Furthermore it is strongly electron-hole asymmetric due to the different spin splitting of valence and conduction bands.

Finally, we note that, although the role of temperature is beyond the scope of this work, spin transport becomes specially interesting in the case of few-layer systems where temperature may drive a crossover from indirect toward direct bandgap regimes, as in the case of MoSe$_2$. In fact, multilayer samples have been shown to effectively behave as single layers, by means of thermally decoupling adjacent sheets via interlayer thermal expansion.\citep{Tongay_etal_2012} This procedure could lead to long spin lifetimes, as the ones needed for spintronic applications, even without the requirement of isolation of single-layer samples. 

\section{Acknowledgements}
We thank E. Cappelluti and F. Guinea for very useful conversations. R. R. acknowledges financial support from the Juan de la Cierva Program (MINECO, Spain). H. O. acknowledges financial support through grant JAE-Pre (CSIC, Spain).

\bibliography{MoS2}
\end{document}